\documentclass[aps,prl,numerical,reprint,prl,showpacs,amsfonts,amssymb,amsmath,mathbbol,twocolumms, letterpaper,groupedaddress,footinbib,floatfix]{revtex4-1}
\usepackage{amsmath}  
\usepackage{graphics}  
\usepackage{color}      
\usepackage{epsfig}
\usepackage{epstopdf} 
\usepackage{bm}        
\usepackage{verbatim}

\usepackage{xspace}
 \newcommand{\ket}[1]{\ensuremath{|#1\rangle}\xspace}

\usepackage{graphicx}
\usepackage{multirow}
\usepackage{rotating}
 
\usepackage{amsfonts}
\usepackage{color}

\usepackage[colorlinks=true, linkcolor=blue,citecolor=blue,urlcolor=blue]{hyperref}
\newcommand{\overbar}[1]{\mkern 1.5mu\overline{\mkern-1.5mu#1\mkern-1.5mu}\mkern 1.5mu}

\begin{document}
\draft \narrowtext \sloppy
\title{ Hund's rule-driven Dzyaloshinskii-Moriya interaction at 3$d$-5$d$ interfaces}

\author{  A.\,Belabbes$^{1}$}
\email {abderrezak.belabbes@kaust.edu.sa}
\author{G.\,Bihlmayer$^{2}$}
\author{ F.\,Bechstedt$^{3}$}
\author{ S.\,Bl\"ugel$^{2}$}
\author{ A.\,Manchon$^{1}$}
\email{aurelien.manchon@kaust.edu.sa}

\affiliation{$^1$
\mbox{King Abdullah University of Science and Technology (KAUST), Physical Science and Engineering Division (PSE)}
\mbox{ Thuwal 23955-6900, Saudi Arabia}}
\affiliation{$^2$
\mbox{ Peter Gr\"unberg Institut and Institute for Advanced Simulation, Forschungszentrum J\"ulich and JARA,  D-52425 J\"ulich }
\mbox{ Germany}}
\affiliation{$^3$ 
\mbox{Institut f\"ur Festk\"orpertheorie und  -optik, Friedrich-Schiller-Universit\"at Jena, Max-Wien-Platz 1, 07743 Jena, Germany}}

\date{\today}
\begin{abstract}
Using relativistic first-principles calculations, we show that the chemical trend of Dzyaloshinskii$-$Moriya interaction (DMI) in 3$d$-5$d$ ultrathin films follows Hund's first rule with a tendency similar to their magnetic moments in either the unsupported 3$d$ monolayers or 3$d$-5$d$ interfaces. We demonstrate that, besides the spin-orbit coupling (SOC) effect in inversion asymmetric noncollinear magnetic systems, the driving force is the 3$d$ orbital occupation and their spin flip/mixing processes with the spin-orbit active 5$d$ states control directly the sign and magnitude of the DMI. The magnetic chirality changes are discussed in the light of the interplay between SOC, Hund's first rule,  and the crystal field splitting of $d$ orbitals.

\end{abstract}
\maketitle

\paragraph{Introduction} Chiral objects are ubiquitous in science \cite{Kondepudi-1990} and pose fundamental challenges such as the importance of chiral molecules in commercial drugs \cite{MAUREEN-2004} or the dominance of matter over antimatter in the universe. Magnetic materials lacking inversion symmetry can host chiral magnets and present a unique platform for the exploration and control of chiral objects. The dynamic development of this field has been recently illustrated by the observation of the magnon Hall effect \cite{Onose-science-2010,manchon-PRB-224403} or the achievement of room temperature magnetic skyrmions \cite{Moreau-Luchaire-2016,Boulle-2016,Jiang283,Tokunaga-2015}, opening avenues for robust high density data storage \cite{Fert-2013}.\par

A crucial ingredient for the generation of such chiral textures is the Dzyaloshinskii-Moriya antisymmetric magnetic interaction (DMI) \cite{dzyaloshinskii, moriya} arising from spin-orbit coupling (SOC) in inversion asymmetric magnets. Originally proposed in the context of Mott insulators \cite{moriya}, weak metallic ferromagnets and spin glasses \cite{fert80}, major attention has been recently drawn toward the nature of DMI at transition-metal (TM) interfaces. Such interfaces, consisting of a stack of 3$d$/5$d$(4$d$) transition metals, have been intensively investigated from the viewpoint of mainstream spintronics resulting in the recent development of spin-orbit torques \cite{miron2011,Luqiao-Liu-2014}  and domain wall-based devices \cite{Yang-Nature-2015}. In these systems, the interfacial DMI gives rise to several exotic magnetic phases such as N\'eel domain walls \cite{Thiaville-EPL-2012, ryu-k-s-nature-2013}, spin spirals \cite{bode2007}, and skyrmions with a defined chirality \cite{heinze2011,Moreau-Luchaire-2016,Boulle-2016,Jiang283}.

The ability to understand and control the sign and strength of DMI remains the big challenge of research in magnetism and may open new approaches to future nanoscale magnetic devices \cite{abdu-arxiv-2016}.
It demands a qualitative description of the physics of DMI that can serve as a guideline for materials and interface design. While such models are available in the context of Mott insulators \cite{moriya}, spin glasses \cite{fert80} and magnetic Rashba gases \cite{Kyoung-Whan-PRL-2013}, such a phenomenology is still lacking for transition-metal interfaces. In fact, the high complexity of interfacial hybridization hinders the development of qualitative and quantitative predictions in these materials combinations. Only few isolated examples have been investigated from first principles \cite{kashid,Bornemann-PRB-2012, SimonJPCM-2014,dupe,PRB-Zimmermann-2016,Schweflinghaus-PRB-2016,PRL-Hongxin2015}. It is therefore crucial to apply such studies and examine the trends of DMI across 3$d$/5$d$ transition metal interfaces, in order to identify the underlying physical mechanisms and develop a predictive physical picture.

In this Letter, we present the first systematic and comprehensive theoretical analysis of DMI for a large series of 3$d$ transition metals (V, Cr, Mn, Fe, Co, Ni) as overlayers on 5$d$-TMs (W, Re, Os, Ir, Pt, Au) substrates. We demonstrate that the sign and magnitude of DMI are directly correlated to the degree of 3$d$-5$d$ orbital hybridization around the Fermi energy, which can be controlled by the intra-atomic Hund's exchange field of the 3$d$ overlayer \cite{Hund1,Khajetoorians-2015}. \par

\paragraph{First-principles method} In order to understand the behavior of DMI in 3$d$-5$d$ ultrathin films we have performed density􀀀 functional theory (DFT) calculations in the local density approximation (LDA) \cite{perd-zunger-LDA} to the exchange correlation functional, using the full 􀀀potential linearized augmented plane wave (FLAPW) method in film geometry \cite{Wimmer1981} as implemented in the  J\"ulich DFT code F{\footnotesize LEUR} \cite{Fleur}.
Both collinear and noncollinear magnetic configurations have been studied employing an asymmetric film consisting of six substrate layers of 5$d$-TM covered by a pseudomorphic
3$d$-TM monolayer on one side of the film at the distance optimized for the energetically lowest collinear magnetic state. 
For the non-collinear calculations we used a $p$(1$\times$1) unit cell applying the generalized Bloch theorem \cite{KurzPRB-2004}. We considered 512 and 1024 \textbf{k}$_{\parallel}-$points in the two-dimensional Brillouin zone (2D$-$BZ) for calculations including the scalar-relativistic effects and SOC treated within first-order perturbation theory, respectively. More details on the computation of the spin-spiral and DMI contribution are given in the Supplementary Material \cite{Supp-Matt}.
 
\begin{figure}[h!]
\includegraphics[width=0.48\textwidth ]{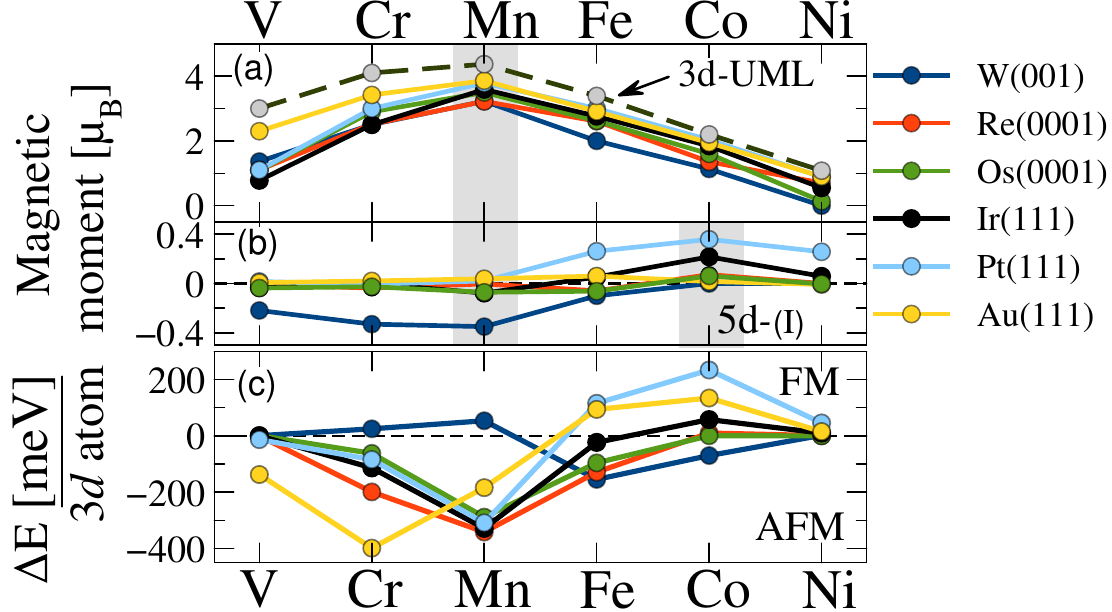}
\caption{(Color online) (a) Calculated magnetic moments of the 3$d$ TM monolayers
on 5$d$ substrates compared to the moments of 3$d$ UML indicated by the dashed black line. (b) The magnetic moments of interface 5$d$ atoms \cite{AFM-order}. (c) The magnetic order of 3$d$ monolayers on 5$d$ substrates using different configurations: FM state, row-wise $p$(1$\times$2)- and checkerboard $c$(2$\times$2)-AFM states for the square lattice (001), FM and AFM for the (111) and (0001) oriented surfaces. Positive $\Delta E= E_{\rm{AFM}}$-$E_{\rm{FM}}$ indicates a FM ground state, while negative values denote an AFM order. 
\label{fig1}}
\end{figure}

\paragraph{Magnetism of 3$d$/5$d$ interfaces}  In \textcolor{blue} {Fig. \ref{fig1}(a)} we display the variation of the magnetic moment across the 3$d$/5$d$ interfaces for their magnetic ground state at equilibrium interlayer distances. A major trend appears: Mn overlayer has the highest magnetic moment, regardless of the substrate, and it gradually decreases for chemical elements on both sides of Mn in the 3$d$ TM row of the periodic table. The magnetic moment, although reduced by about 1-3 $\mu_{\rm{B}}$, qualitatively follows the total spin S of the 3$d$ shell. This trend reveals that the intra-atomic exchange is controlled by the atomic-like nature of the orbitals according to Hund's first rule \cite{StefanPRL-1992,FerrianiPRB-2005,Hund1}. We note that the local magnetic moment of the 3$d$/5$d$ films (solid lines) is reduced with respect the magnetic moment of the (ideal) unsupported monoatomic 3$d$ layers (UML - dashed line). This is a consequence of the increased bonding at the interface due to the orbital hybridization of surface and overlayer states [see \textcolor{blue} {Fig. \ref{fig1}(a)}]. 

\textcolor{blue} {Fig. \ref{fig1}(b)} shows that the substrate exhibits an induced spin polarization since the increased bonding with 3$d$ overlayer enhances the magnetic moment of the neighboring 5$d$ atoms at the interface. From this figure we conclude that W, Pt and to some extent Ir are highly polarizable substrates, while the other substrates considered show a relatively small induced magnetic moment at the interface. The induced magnetic moment for Pt and W substrates couples ferromagnetically and antiferromagnetically with the 3$d$ overlayer, respectively. Their strong local-spin polarizability is mainly due to the high spin susceptibility originating from the large Stoner exchange parameter. For the Au substrate, on the other hand, there is almost no or only very weak polarization since the 3$d$ orbitals do not hybridize with
 the energetically low-lying 5$d$ states of Au.  \par

To complete the description of the transition-metal interfaces, we analyze the magnetic stability of 3$d$ overlayers on various 5$d$ substrates in terms of the total energy difference $\Delta E= E_{\rm{AFM}}$-$E_{\rm{FM}}$ as shown in \textcolor{blue} {Fig. \ref{fig1}(c)}. We primarily focus on three collinear configurations: the FM state, row-wise $p$(1$\times$2)-AFM state for (111) and (0001) surfaces and the checkerboard $c$(2$\times$2)-AFM states for (001) oriented surfaces (see \textcolor{blue} {Fig. \ref{fig1}} in the Supplementary Material \cite{Supp-Matt}).
 For a W(001) substrate, we clearly observe the opposite trend for the early 3$d$ overlayer elements V, Cr, and Mn compared to all other considered 5$d$ substrates, in good agreement with the theoretical study by Ferriani $et$ $al$. \cite{FerrianiPRB-2005}, who predicted the same energetic ordering for 3$d$/W(001). We find that the magnetic ground state is AFM for all substrates, while it is FM for W with small energy differences between the two magnetic configurations. In the case of Mn/W(001) the row-wise $p$(1$\times$2)-AFM state is energetically more favorable than the checkerboard $c$(2$\times$2)-AFM configuration by $\sim 0.3$ eV/Mn atom. However, for Ni, Co, and Fe, moving from right to left through the 5$d$ elements, we observe a strong tendency from FM ordering toward an AFM coupling as a function of the 5$d$ band filling of the substrate. 

We notice that the 3$d$-5$d$ interface states around the Fermi energy and their relative lineup control the competition between FM and AFM coupling of the deposited 3$d$ atoms \cite{FerrianiPRB-2005,Bechstedt:2003:Book:PSP1,StefanblugelC1}.
This especially holds when the nearest-neighbor exchange interaction $\Delta E= E_{\rm{AFM}}$-$E_{\rm{FM}}$ is small as shown in \textcolor{blue} {Fig. \ref{fig1}(c)}. In this case, complex magnetic textures can be expected in the presence of SOC since the antisymmetric exchange DM interaction \cite{dzyaloshinskii, moriya} contributes considerably to the total energy. Indeed, if it is sufficiently strong to compete with the magnetocrystalline anisotropy and the Heisenberg exchange, it can stabilize long-range chiral magnetic order such as skyrmions or homochiral spin spirals \cite{bode2007,heinze2011}.\par

\paragraph{Dzyaloshinskii-Moriya interaction} Phenomenologically, the DMI has the typical form $E_{\mathrm{DM}}=\sum_{i,j} \mathrm{\bold{D}_{ij}} \cdot (\bold{S}_{i} \times \bold{S}_{j})$, where $\mathrm{\bold{D}_{ij}}$ determines the strength and sign of DMI, and $\bold{S}_{i}$ and $\bold{S}_{j}$ are magnetic spin moments located on neighboring atomic sites $i$ and $j$ (see Supplementary Material for more details \cite{Supp-Matt}). The energy contribution to DMI due to SOC treated in first order perturbation theory corresponds to the sum of all energy shifts from filled states, $ E_{\mathrm{DMI}}(\textbf{q})= \sum_{\mathrm{k}\nu}^{\mathrm{occ.}} n_{\mathrm{k}\nu}(\textbf{q})  \delta \epsilon_{\mathrm{k}\nu}(\textbf{q})$, with $n_{\mathrm{k}\nu}(\textbf{q})$ as the occupation numbers of state $\ket{\psi_{\textbf{k}\nu}(\textbf{q})}$ ($\nu$-band index, $\textbf{k}$-Bloch vector) and $\textbf{q}$ as the wave-vector propagation of the spin-spiral. Here, the energy shift of the occupied states with respect to the scalar-relativistic (SR) calculation corresponds to $\delta \epsilon_{\mathrm{k}\nu}=\epsilon_{\mathrm{k}\nu}^{\mathrm{SOC}}-\epsilon_{\mathrm{k}\nu}^{\mathrm{SR}}$. In the limit of smooth magnetic textures ($\textbf{q} \rightarrow 0$), the DMI can be directly determined by a linear fit $E_{\mathrm{DMI}}(\textbf{q}) \approx Dq$ \cite{HeidePRB:2008}. In order to further understand the layer-resolved DMI energy $E_{\mathrm{DMI}}^{\mu}(\textbf{q})$ ($\mu$ labels the atom in the unit cell), we consider the site decomposition of the SOC operator $\mathcal{H}_{\mathrm{so}}=\Sigma_{\mu} \xi (r^{\mu})  \sigma \cdot \mathrm{L}^{\mu}$, where $\xi $ is the SOC strength related to the spherical muffin-tin potential $V(r^{\mu})$, $\xi \sim r^{-1} \mathrm{d}V/\mathrm{d}r$, $\textbf{r}^{\mu}=\textbf{r}-\textbf{R}^{\mu}$ and $|\textbf{r}^{\mu}|<R_{\mathrm{MT}}^{\mu}$. $\textbf{R}^{\mu}$ references the center and $R_{\mathrm{MT}}^{\mu}$ is the radius of the $\mu$th muffin-tin sphere, and $\mu$ runs over all atoms in the unit cell. \par

\begin{figure}[t]
\includegraphics[width=0.48\textwidth ]{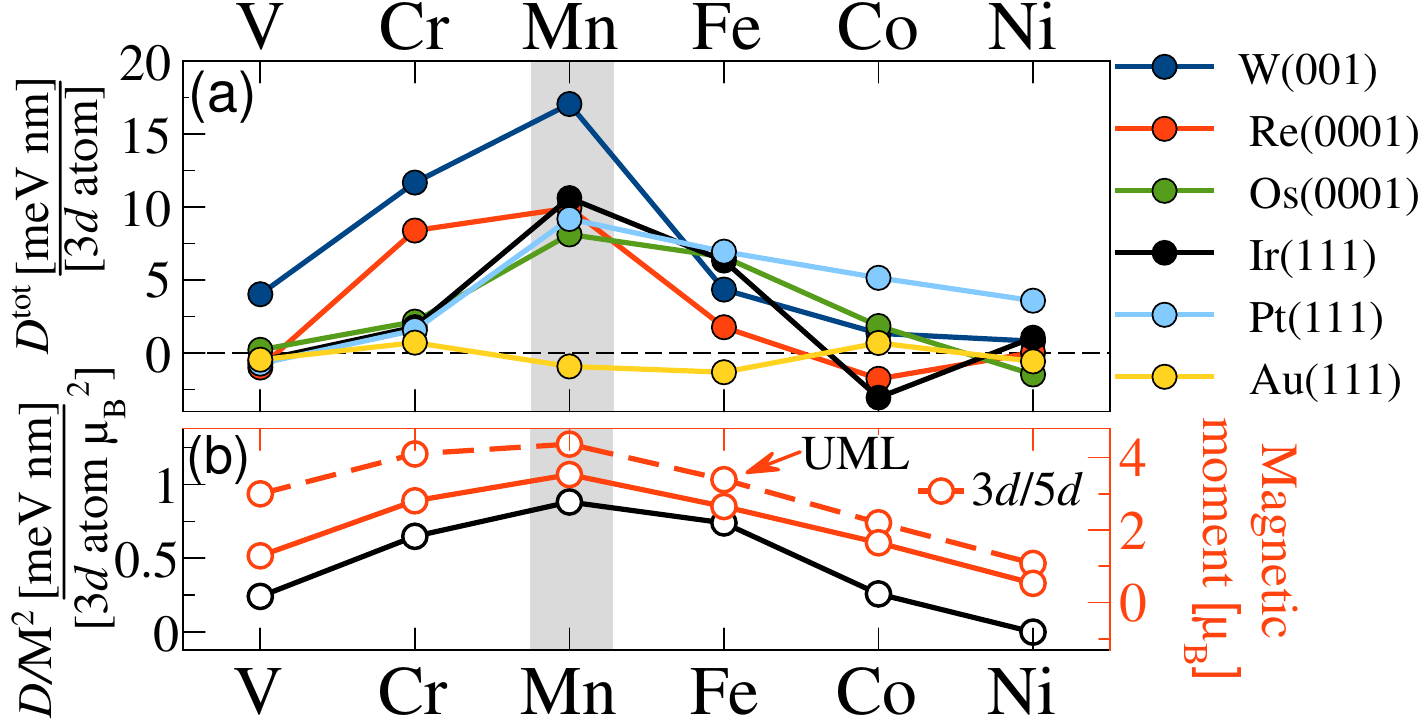}
\caption{(Color online) (a) Strength and sign of Dzyaloshinskii$-$Moriya interaction (DMI) $D^{\mathrm{tot}}$ in 3$d$ TM monolayers on 5$d$ substrates calculated around their magnetic ground state combining the relativistic SOC effect with spin spirals. A positive sign of $D^{\mathrm{tot}}$ indicates a left-rotational sense or "left chirality". (b)  correlation between $E_{\mathrm{DMI}}$ $\sim$ $D^{\mathrm{tot}}/M^{2}_{3d/5d}$ averaged over 3$d$/5$d$ interfaces (black line) versus the adlayer, the magnetic moments in 3$d$ TM UML (dashed red line), and the local magnetic moment per atom averaged over 3$d$/5$d$ interfaces (solid red line).
\label{fig2}}
\end{figure}

The central result of this Letter is summarized in \textcolor{blue} {Fig. \ref{fig2}(a)} (see also Tables I and II in \cite{Supp-Matt}). There the total DMI energy $D^{\mathrm{tot}}$ is represented as a function of the 3$d$ overlayer element for various 5$d$ substrates. Apart from 3$d$/Au(111) interfaces, the calculations reveal a very surprising trend in which the modulus of the $\textit{D}$-vector, i.e., the DMI energy divided by the square of the spin magnetic moment $M^{2}_{3d/5d}$, across the 3$d$/5$d$ interfaces follows Hund's first rule  with a tendency similar to their magnetic trends in either the 3$d$ UML or 3$d$/5$d$ ultrathin films [see \textcolor{blue} {Fig. \ref{fig2}(b)}]. In low dimensional systems the spin moments as function of the number of $d$ electrons are well described by Hund's first rule \cite{StefanPRL-1992,Hund1,MediciPRL-2011,StefanblugelC1}. For the $\textit{D}$-vector, such a correlation has been neither experimentally nor theoretically demonstrated for 3$d$/5$d$ thin films. This fact is surprising as it is opposite to what was expected from the knowledge of magnetism in bulk and thin films, especially in view that such a correlation does not hold for the proximity induced magnetization, Heisenberg exchange parameters, $J_{ij}$, and the magnetic energy anisotropy \cite{abdu-arxiv-2016,PRLGayles-2015,StefanblugelC1,Ryu-Natcom-2014,Kvashnin-PRL-2016} [see also \textcolor{blue} {Fig. \ref{fig1}(a-b-c)} and {Fig. \ref{fig2}}]. Indeed, the nearest-neighbor exchange interaction ($\bold{\emph{J}}_1$) for 3$d$ TMs follows perfectly the Bethe-Slater curve and not Hund's first rule \cite{Kvashnin-PRL-2016}.  More specifically, since DMI emerges from a complex interplay between $(i)$ degree of spin-polarization of 3$d$/$5d$ interface atoms and their band filling, $(ii)$ strength of SOC in the underlying heavy metal 5$d$-substrate, and $(iii)$ the inversion symmetry breaking at the interface, one does not necessarily expect a direct correlation between the magnetism of the 3$d$ overlayer and the DMI. In the discussion part we will explain in more detail the physical reasons behind the unexpected trend. \par
%
%Here, we argue that, naturally the energy-terms resulting from these interactions are proportional to $M^2$, but also the interaction parameters, $D^{\mathrm{tot}}/M^2$, follow the behavior of M due to their special dependence on the 3$d$/5$d$ band-alignment. 

In \textcolor{blue} {Fig. \ref{fig2}(a)} the largest absolute DMI values are obtained for Mn/5$d$ films, regardless of the substrate, with a maximum value of 17 meV nm for Mn/W(001). They monotonically decrease toward V and Ni atoms. 
%The strong DMI in the Mn/W(001) system seems unusual at first glance but can be understood on the basis of the surface coordination number ($C_{4v}$ symmetry with four W nearest neighbors \cite{FerrianiPRB-2005,feriani-PRL-2008}), the bandwidth of the partially filled 5$d$ states around the Fermi energy, strong SOC, and the large magnetic moment as mentioned above [$\sim$ 3.22 $\mu_{B}$ for Mn and the induced W moment at the interface $\sim$ 0.35 $\mu_{B}$, see \textcolor{blue} {Fig. \ref{fig1}(a)(b)}].
 In contrast, despite the large SOC of Au, DMI in 3$d$/Au almost vanishes due to the completely filled $d$ shell of the Au substrate, irrespective of the 3$d$ overlayer [see \textcolor{blue} {Fig. \ref{fig2}(a)}]. This remarkable finding demonstrates that the DMI does depend critically not only on SOC and lack of the inversion symmetry, but also on the $d$ wavefunction hybridization of the studied 3$d$/5$d$ interface. The latter affects significantly the interlayer hopping of electrons and, consequently, the magnetic coupling between the 3$d$ overlayer. It is also worthwhile to note that most of 3$d/5d$ interfaces have a positive sign of DMI, left- or right-rotating depending on their magnetic ground state [cf. \textcolor{blue} {Fig. \ref{fig1}(c)} and \textcolor{blue} {Fig. \ref{fig2}}]. 

\begin{figure}[h]
\includegraphics[width=0.5\textwidth ]{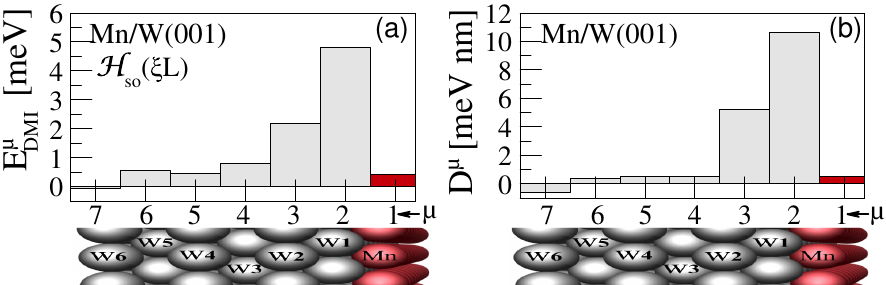}
\caption{(Color online) (a) Layer-resolved $E_{\mathrm{DMI}}^{\mu}(\textbf{q})$ to the energies of long-period lengths for Mn/W(001) and (b) the DMI-strength $D^{\mu}$ is extracted through a linear fit [$E_{\mathrm{DMI}}^{\mu}(\textbf{q}) \approx D^{\mu}q$]. Note, the positive sign of $D^{\mathrm{tot}}$ ($D^{\mathrm{tot}}=\Sigma_{\mu} D^{\mu}$) indicates a left chirality.
\label{fig3}}
\end{figure}

\paragraph{Discussion} 
For the analysis below,  it is worth emphasizing that the delocalized 5$d$ wave functions are responsible for the SOC matrix elements $\mathcal{H}_{\mathrm{so}}(\xi \mathrm{L})$ and make essential contributions to DMI [see \textcolor{blue}{Fig. \ref{fig3}(a)}]. This behavior is confirmed by the layer-resolved DMI parameter $D^{\mu}$ [\textcolor{blue}{Fig. \ref{fig3}(b)}], which indicates that the sign and strength of DMI are mainly ascribable to the large contribution of the 5$d$ surface: 80\%  of the total DMI strength $D^{\mathrm{tot}}$ comes from the first two 5$d$ surface layers depending only weakly on the $3d􀀀$ overlayer. An analogous behavior has been identified for the Rashba effect, where 90\%  of the Rashba splitting is dominated by 5$d$ surface state wavefunction \cite{sergiy-PRB-2016,Bihlmayer20063888}. However, despite the weak SOC in the 3$d$ overlayer their intra-atomic exchange field can easily modify the electronic structure around the Fermi energy and consequently change the strength of DMI. \par

\
\par
The results displayed in \textcolor{blue} {Figs. \ref{fig1}} and \textcolor{blue} {\ref{fig2}} demonstrate a systematic correlation between DMI and magnetic moment on the one hand, and DMI and the energetic positions of 3$d$/5$d$ states on the other hand. This mechanism can be understood by examining the band-alignment of the 3$d$ and 5$d$ states and their spin flip/mixing processes, since the anisotropic exchange mechanism requires spin-flip transitions between occupied and unoccupied states that involve spin-orbit active states \cite{moriya,kashid}. Note that the intermediate 5$d$ states
are necessary for spin-flip process to unoccupied states of the other spin channel. \par
\begin{figure}[t]
\includegraphics[width=0.46\textwidth]{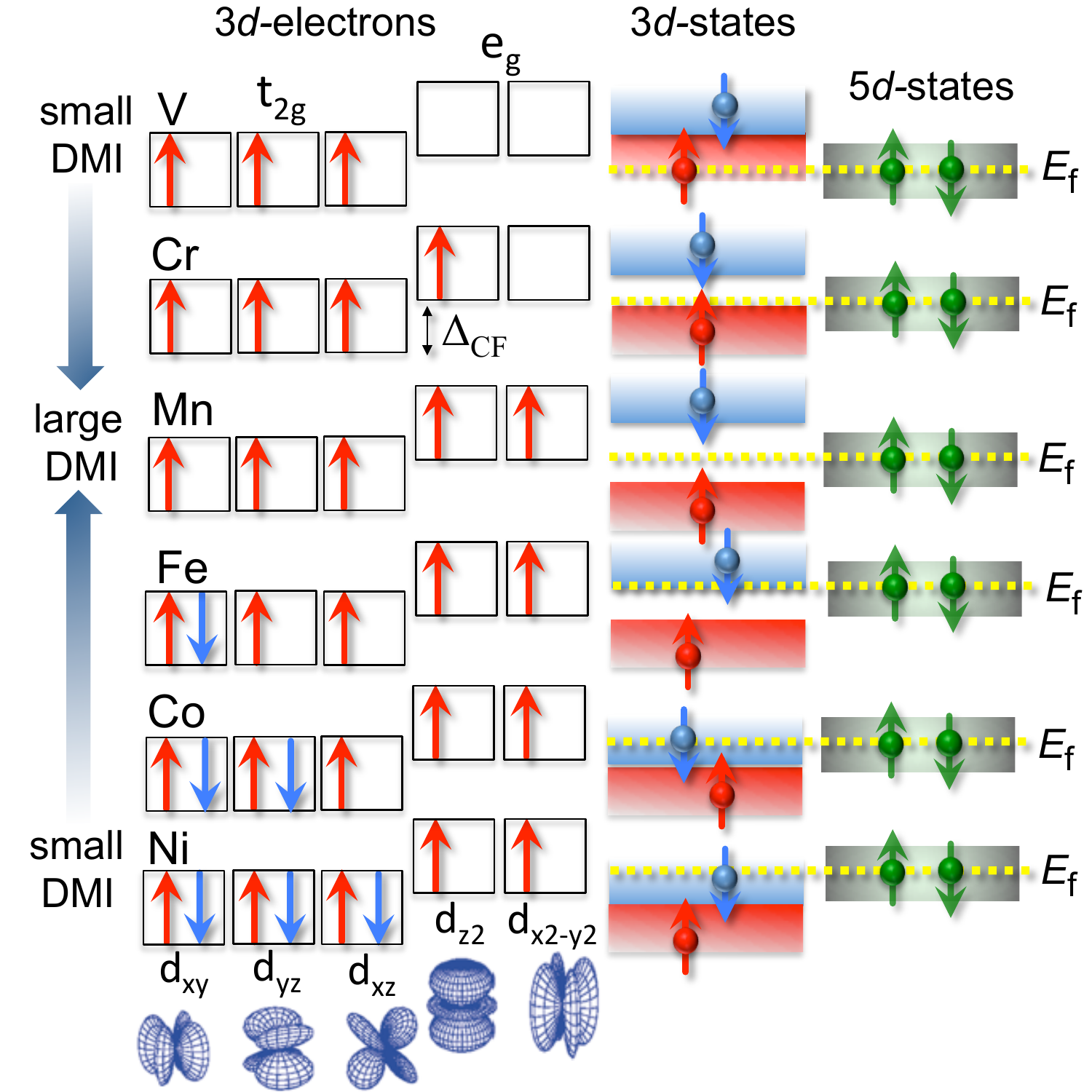}
\caption{(Color online) Left side: filling with electrons of 3$d$ TMs elements into the five 3$d$-orbitals according to Hund's first rule, spin-up  and -down are shown by red and blue arrows, respectively. On the right side we show the spin-split band positions of 3$d$ states with respect to 5$d$-W states. Note, since the 5$d$ bandwidth is significantly larger than the crystal field splitting the 5$d$ states are degenerate at the Fermi level.  
$\Delta_{\mathrm{CF}}$ indicates the crystal-field splitting between the $t_{2\mathrm{g}}$ and $e_{\mathrm{g}}$ shells.
%$\Delta_{\mathrm{CF}}$ indicates the crystal-field splitting between the $t_{2\mathrm{g}}$ and
%$e_{\mathrm{g}}$ shells.
\label{fig4}}
\end{figure}

The basic idea is illustrated in \textcolor{blue} {Fig. \ref{fig4}}, where the electronic configuration of 3$d$ orbitals and their spin-split band positions with respect to 5$d$-W  states are displayed. Since the 5$d$ bandwidth is significantly larger than the crystal field splitting and the 5$d$ states are weakly polarized (degenerate and partially filled), the overall physics is mostly governed by the band lineup of 3$d$ spin channels, themselves determined by Hund's first rule. According to this rule, for the V and Ni overlayers both spin channels are almost occupied or unocupied and consequently transitions between these states do not contribute anymore to the DMI. We emphasize that the occupied and unoccupied states should be available for the 3$d$ electrons to allow for spin-flip excitations. In the case of Co, some spin-down states become unoccupied and transitions into these states contribute only weakly to the DMI.  However, in the case of Mn the filling of the five Mn-3$d$ orbitals adopts a stable \textit{"high spin state"} due to the small crystal-field splitting between the $t_{2\mathrm{g}}$ and $e_{\mathrm{g}}$ shells [see \textcolor{blue}{Fig. \ref{fig4}}]. As a result, the spin-up (spin-down) channels are entirely occupied (unoccupied) and all transitions contribute to DMI through the intermediate spin-orbit active 5$d$ states. In other words, the 3$d$-5$d$-3$d$ electron hopping is facilitated, resulting in a large DMI. Note that the situation is almost similar for the half-filled Fe and Cr atoms but the \textit{exchange splitting} is reduced where most of the Fe spin-down (Cr spin-up) states are still unoccupied (occupied). This fact clearly explains the sensitivity of DMI on the choice of 3$d$ overlayer in \textcolor{blue} {Fig. \ref{fig2}}. Since the electronic configuration of the 5$d$ orbitals is governed by either the bandwidth or crystal field splitting, these scenarios are valid for all 3$d/$5$d$ interfaces, irrespective of the substrate. The Au substrate is an exception since the spin-orbit 5$d$ states should dominate at the Fermi level and have to be energetically close to unoccupied minority spin states to facilitate the spin-flip processes necessary for the DMI.\par  

 Based on the above scenarios, the interplay between Hund's exchange, crystal field splitting, and SOC should be generally considered in any \textit{"design"} of the DMI. The band-lineup determined by Hund's rule controls not only the 3$d$ spin-flip transitions states but also their spin-mixing processes with the spin-orbit active 5$d$ states [see \textcolor{blue} {Fig. \ref{fig2}} and \textcolor{blue} {Fig. \ref{fig4}}]. Summarizing the results for all 3$d$/5$d$ interfaces, we conclude that for an interface with 3$d$ overlayer that has a band gap around the Fermi energy in both spin channels, a rather large DMI energy should be expected. This fact also relates to the gradual decrease of DMI for chemical elements on both sides of Mn in the 3$d$ TM row of the periodic table, although in low-symmetry environments additional factors may play a role \cite{Schweflinghaus-PRB-2016}. In this context, atomic Hund's first rule is a powerful guideline to control the sign and magnitude of DMI in particular since the total energy contribution also involves a proportionality to $M^{2}_{3d/5d}$. \par
 
%
% (\textcolor{blue}{Fig. \ref{fig5}}), and hybridization occurs depending on the orbital filling. As a consequence the degree of 3$d$/5$d$ hybridization reduces gradually the DMI energy for 3$d$ elements on both side of Mn in the 3$d$ TM row of the periodic table. In this case, the sign and magnitude of DMI depend on the shifts of the single-particle energies with respect to the scalar-relativistic eigenvalues due to SOC ($\delta \epsilon_{\mathrm{k}\nu}=  \epsilon_{\mathrm{k}\nu}^{\mathrm{SOC}}-\epsilon_{\mathrm{k}\nu}^{\mathrm{SR}}$) in the energy range around the Fermi level where 3$d$ states hybridize with the 5$d$ states of the substrate \cite{kashid}. \par

%However, in the case of Mn/W(001) the Mn spin-majority $d$ band is nearly filled and the spin-minority $d$ band is nearly empty. The filling of the five Mn-3$d$ orbitals adopts a stable \textit{"high spin state"} due to the small crystal-field splitting $\Delta_{\mathrm{CF}}$ between the $t_{2\mathrm{g}}$ and $e_{\mathrm{g}}$ shells [see \textcolor{blue}{Fig. \ref{fig5}}]. In contrast, W has degenerate, partially filled 5$d$ states due to the strong $\Delta_{\mathrm{CF}}$. The hybridization between Mn 3$d$ states and W 5$d$ states is weak resulting in a large interlayer distance in comparison with other 3$d$/W(001) interfaces \cite{FerrianiPRB-2005}. As a consequence, the large asymmetry of the 5$d$ surface state wavefunction remains essentially unmodified, which contributes largely to DMI in the presence of SOC. 

In summary, we have predicted the systematic trend of the DMI in 3$d$-5$d$ ultrathin films using first-principles calculations. In particular, we demonstrated that the sign and strength of DMI depend strongly on degree of hybridization between 3$d$-5$d$ states around the Fermi level. Furthermore, in addition to $(i)$ strength of SOC in the underlying heavy metal 5$d$-substrate, $(ii)$ the degree of the inversion symmetry breaking at the interface and $(iii)$ 5$d$ band filling, we show that the  driving force behind the peculiar behavior of DMI is the 3$d$/5$d$ band-lineup controlled by the Hund's rule filling of $3d$ shells, which also plays a decisive role in the general picture of spin dynamics. We anticipate that our prediction will provide guidance for the experimental realization and further investigation of chiral properties of ultra-thin magnetic films.\par

A.B. and A.M. acknowledge financial support from the King Abdullah University of Science and Technology (KAUST) through the Award No OSR-CRG URF/1/2285-01 from the Office of Sponsored Research (OSR). We acknowledge computing time on the supercomputers SHAHEEN, NOOR, and SMC at KAUST Supercomputing Centre and JUROPA at the J\"ulich Supercomputing Centre (JSC). 

\pagebreak
\widetext
\begin{center}
\textbf{\large Supplemental Materials: Hund's rule-driven Dzyaloshinskii-Moriya interaction at 3$d$-5$d$ interfaces}
\end{center}
\subsection{Spin-spiral and Dzyaloshinskii-Moriya Interaction (DMI)}

\begin{figure}[h]
\includegraphics[width=0.55\textwidth ]{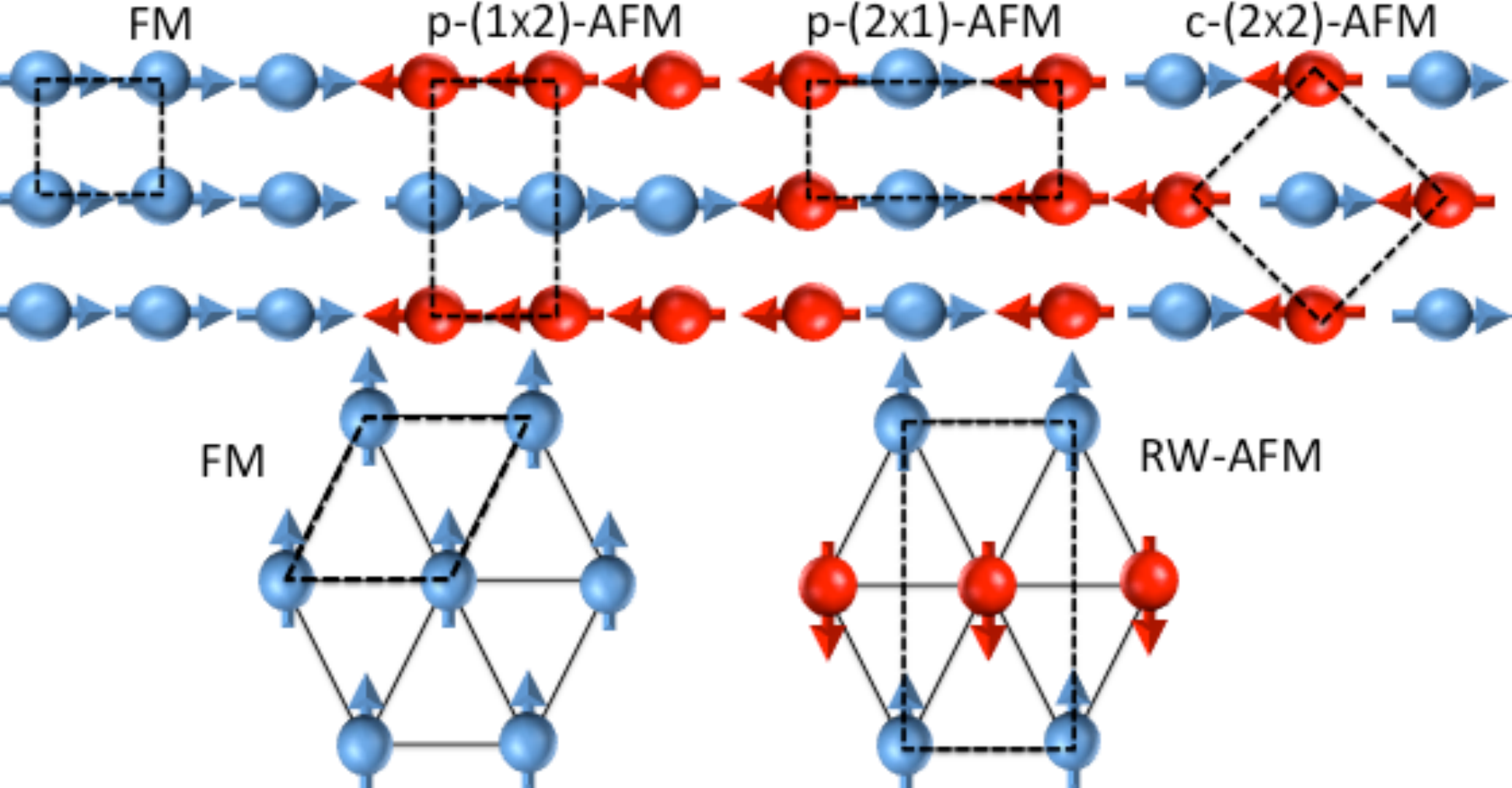}
\caption{(Color online) Unit cell sketch of the investigated magnetic configurations of 3$d$-TMs on 5$d$ substrates: FM state, row-wise $p$(1$\times$2)- or $p$(2$\times$1)- and checkerboard $c$(2$\times$2)-AFM states for the square lattice (001) (upper row), FM and row-wise AFM for the (111) and (0001) oriented surfaces (lower row). For the substrates the ground-state crystal structures bcc (W), hexagonal (Re, Os), and fcc (Ir, Pt, Au) with corresponding surface orientations (001) (W), (0001) (Re, Os), and (111) (Ir, Pt, Au) are investigated.
\label{fig5}}
\end{figure}

In order to investigate the DMI, first we self-consistently calculate the total energy of homogeneous magnetic spin-spirals employing the generalized Bloch theorem within the scalar-relativistic approach \cite{KurzPRB-2004}. We have considered the energy dispersion $E(\textbf{q} )$ of planar spin spirals which are the general solution of the Heisenberg Hamiltonian, i.e., states in which the magnetic moment of an atom site $\textbf{R}_{i}$ is given by $\textbf{M}_{i}=M [\mathrm{cos}(\textbf{q} \cdot \textbf{R}_{i}), \mathrm{sin}(\textbf{q} \cdot \textbf{R}_{i}), 0]$ where $\textbf{q}$ is the wave vector propagation of the spin spiral. By imposing the N\'eel spin spirals along the high-symmetry lines of 2D-BZ in either square or triangular lattices, we can scan all possible magnetic configurations that are described by a single $\textbf{q}$ vector. So, varying the $\textbf{q}$ vector with small steps along the paths connecting the high-symmetry points, we find the well-defined magnetic phases, for hexagonal lattices: FM state at $\overbar{\Gamma}$-point ($\textbf{q}$=0), RW-AFM state at the $\overbar{M}$-point, and periodic 120$^{\circ}$ N\'eel state at the $\overbar{K}$-point, while for square lattice, we find FM state at $\overbar{\Gamma}$-point, RW $p$(1$\times$2)-AFM state at the $\overbar{X}$-point, and $\overbar{M}$-point characterizes the checkerboard $c$(2$\times$2)-AFM (see \textcolor{blue} {Fig. \ref{fig5}}).  When the energy $E(\textbf{q})$ along the high-symmetry lines of 2D-BZ is lower than any of the collinear magnetic phases studied previously, the system most likely adopts an incommensurate spin􀀀-spiral magnetic ground􀀀 state structure. \par

\begin{figure}[t]
\includegraphics[width=0.5\textwidth ]{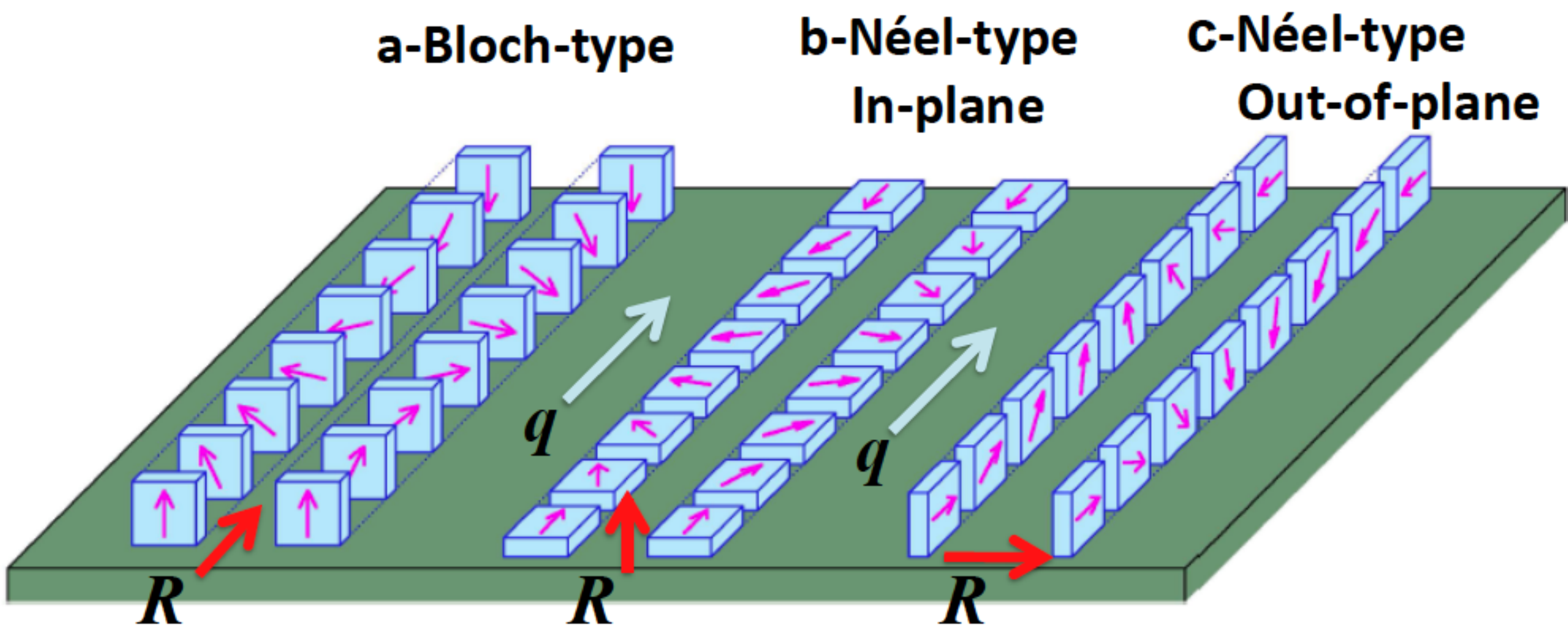}
\caption{(Color online) Schematic representation of spin$-$spirals with different propagation directions $\textbf{q}$ and spin$-$rotation axis $\bf{R}$ (adapted from \cite{Heide-psik}). (b,c) N\'eel$-$type in$-$plane and out$-$of$-$plane $(\bf{R \perp q})$, respectively and (a) Bloch$-$type with $(\bf{R \parallel q)}$. Note that the DMI vanishes for N\'eel$-$in$-$plane and Bloch$-$type due to symmetry arguments.
\label{fig6}}
\end{figure}

 In a second step, we evaluate the DMI contribution from the energy dispersion of spin-spirals by applying the spin-orbit coupling (SOC) treated within first-order perturbation theory combined with the micromagnetic model \cite{HeidePRB:2008, Heide-phy-B}. Phenomenologically, the antisymmetric exchange interaction DMI has the typical form $E_{\mathrm{DM}}=\sum_{i,j} \mathrm{\bold{D}_{ij}} \cdot (\bold{S}_{i} \times \bold{S}_{j})$, where $\mathrm{\bold{D}_{ij}}$ is the DM vector which determines the strength and sign of DMI, and $\bold{S}_{j}$ and $\bold{S}_{j}$ are magnetic spin moments located on neighboring atomic sites $i$ and $j$. Considering  the N\'eel􀀀-type out-of-plane configuration the $\mathrm{\bold{D}_{ij}}$ vector should be oriented in plane and normal to the $\textbf{q}$ vector \textcolor{blue} {Fig. \ref{fig6}(c)}. Note that the DMI term must vanish for both configurations, N\'eel􀀀-type in-plane and Bloch-type spin spirals, due to symmetry arguments \cite{HeidePRB:2008} [see \textcolor{blue} {Fig. \ref{fig6}(a)(b)}]. According to our definition, the vector chirality is characterized by $\mathrm{\textbf{C}}= C \textbf{\^c}=\mathrm{S}_{i}\times \mathrm{S}_{i+1}$, where the direction of the vector spin chirality $\textbf{\^c}$ is considered as spin rotation axis. Thus, left-handed (right-handed) spin spirals correspond to $C$ = +1 ($C$ = -1). 

\begin{figure}[h]
\includegraphics[width=1\textwidth ]{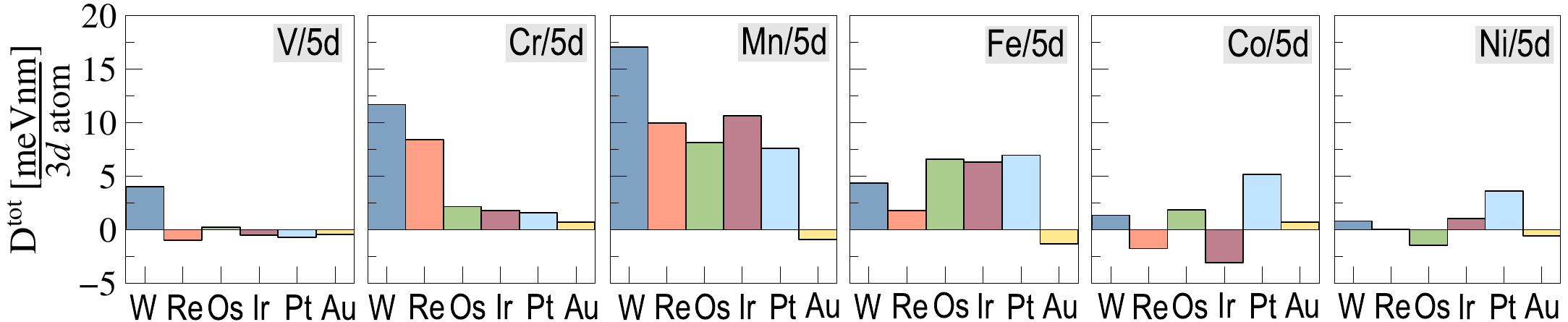}
\caption{(Color online) Strength and sign of Dzyaloshinskii$-$Moriya interaction $D^{\mathrm{tot}}$ in 3$d$ TM monolayers on 5$d$ substrates calculated around their magnetic ground state using the combination of the relativistic effect SOC with the spin spirals.  A positive sign of $D^{\mathrm{tot}}$ indicates a left-rotational sense or "left chirality".
\label{fig7}}
\end{figure}

\begin{table}[bt!]
\begin{ruledtabular}
\caption{\label{tab1}
 Calculated magnetic moments of the 3$d$ TM monolayers on 5$d$ substrate and interface 5$d$(I) atoms at their ground state.
    }
    \begin{tabular}{cccccccccccccccccccccc}

         &   \multicolumn{7}{c}{$\mu_{\mathrm{mag}}$-3$d$ [$\mu_{\mathrm{B}}]$}& \multicolumn{7}{c}{$\mu_{\mathrm{mag}}$-5$d$(I)[$\mu_{\mathrm{B}}]$}& &  \\ \cline{3-8} \cline{10-15}\
  TM&& W(001)&Re(0001)&Os(0001)&Ir(111)&Pt(111)&Au(111)&& W(001)&Re(0001)&Os(0001)&Ir(111)&Pt(111)&Au(111)&
   \\ 
 \hline

V&& 1.36 & 1.09 & 1.10 & 0.78 & 1.10  &2.30 & &       $-$0.22 & $-$0.01 & $-$0.04 & $-$0.03 & 0.01 &0.00 &	\\
 Cr&& 2.50 & 2.90 & 2.52 & 2.50 & 3.00 & 3.42 &&	$-$0.33 & $-$0.03 & $-$0.03 & { }{ }0.00 & 0.00 & 0.02  &	\\
 Mn&& 3.22 & 3.49 & 3.22 & 3.58 & 3.79 & 3.86 &&	$-$0.35 & $-$0.00 & $-$0.07 & $-$0.08 & 0.02 & 0.04 &	\\
 Fe&& 2.00 & 2.62 & 2.61 & 2.76 & 3.00 & 2.91&&	$-$0.10 & $-$0.06 & $-$0.06 &{ }{ }0.05 & 0.26 & 0.06 &	\\
 Co&& 1.13 & 1.60 & 1.36 & 1.84 & 2.03 &1.94	&&{ }{ }0.00 & { }{ }0.07 &{ }{ }0.06 & { }{ }0.21 & 0.36 & 0.02 &  \\
 Ni&& 0.01 & 0.13 & 0.70 & 0.55 & 0.89 &0.90	&&      { }{ }0.00 &{ }{ }0.00 &{ }{ }0.00 & { }{ }0.06 & 0.25 &0.00  & \\

\end{tabular}
\end{ruledtabular}      
\begin{ruledtabular}
\caption{\label{tab1}
Strength and sign of Dzyaloshinskii$-$Moriya interaction $D^{\mathrm{tot}}$ in 3$d$ TM monolayers on 5$d$ substrates calculated around their magnetic ground state using the combination of the relativistic effect SOC with the spin spirals.  A positive sign of $D^{\mathrm{tot}}$ indicates a left-rotational sense or "left chirality". The total energy differences $\Delta E$, positive $\Delta E= E_{\rm{AFM}}$-$E_{\rm{FM}}$ indicates a FM ground state, while negative values denote AFM order.
    }
    \begin{tabular}{cccccccccccccccccccc}

        &  \multicolumn{7}{c}{$D^{\mathrm{tot}}$ [meV nm/3$d$ atom]} &\multicolumn{7}{c}{$\Delta E$ [meV/3$d$ atom]}&  \\ \cline{3-8} \cline{10-15}\newline
  TM&& W(001)&Re(0001)&Os(0001)&Ir(111)&Pt(111)&Au(111)&& W(001)&Re(0001)&Os(0001)&Ir(111)&Pt(111)&Au(111)&
   \\ 
 \hline

V&& { }4.03 & $-$0.99 & { } 0.22 & $-$0.51  & $-$0.73 & $-$0.46    && {   } {  } {  } 0.7 & {}{  }{ } $-$0.6 & {   } {  } {  }0.9 & {   } {  } {  }0.1 & { }$-$13.3 & $-$137.3&\\
Cr&&11.68 &  { } 8.39 &  { } 2.15 &  { } 1.78  &  { } 1.57 &  { } 0.71               &  &   { }  { }{ }25.1 & $-$199.1 &{   }{  }$-$62.2 & $-$113.7 &{ }$-$84.3 &$-$398.8&\\
Mn&&17.06 & { } 9.94 &  { } 8.12 &  { }10.62  &  { } 9.18 & $-$0.92&         &  { }  { }{ }53.6 & $-$341.7 & $-$288.6 & $-$328.7 & $-$309.1 &$-$184.0&\\
Fe&& { }4.35 &  { } 1.77 &  { } 6.59 & { }  6.32 &  { } 6.94  & $-$1.32   &         &   $-$154.6 & $-$128.8 & {   }{  }$-$95.9 &{ } $-$23.2 &{ }{ } 115.7 & { } { }{ }94.0&\\
Co&& { }1.35 & $-$1.77 &  { } 1.85 & $-$3.05  & { }  5.15 &  { } 0.69 &    &        { }$-$70.0 & {   } {  }{  }10.0 &{   } {  } {  }  0.3 & {   } {  }{ }57.8 &{ }{ } 234.0 &{ }{ }{ }134.5 &\\
Ni&& { }0.81 & { } 0.03 & $-$1.47 &  { } 1.06 &  { } 3.58 & $-$0.57 &     &       { }  { } { }2.0 & {   } {  } {  } 0.7 & {   } {  }{  }$-$0.1 &{   } {  } { }9.8 &{ } { }{ } 45.8 &{ } { }{ }15.3&\\
\end{tabular}
\end{ruledtabular}
\end{table}

\bibliography{literatur}

\begin{thebibliography}{49}
\bibitem [{AFM()}]{AFM-order}%
  \BibitemOpen
  \href@noop {} {}\bibinfo {note} {The AFM order does not show any induced spin
  polarization at the interface layer for symmetry reasons}\BibitemShut
  {NoStop}%
\end{thebibliography}
\end{document}